\documentclass[11pt]{article}

\usepackage{fullpage,amsmath,amsfonts,latexsym}
\usepackage{epsfig}
\usepackage{hyperref}
\usepackage{url}
\usepackage{amsthm}
\newcommand\beq{\begin{equation}}
\newcommand\eeq{\end{equation}}
\newcommand\bea{\begin{eqnarray}}
\newcommand\eea{\end{eqnarray}}

\newcommand{\ket}[1]{| #1 \rangle}
\newcommand{\bra}[1]{\langle #1 |}
\newcommand{\braket}[2]{\langle #1 | #2 \rangle}
\newcommand{\proj}[1]{| #1\rangle\!\langle #1 |}
\newcommand{\ba}{\begin{array}}
\newcommand{\ea}{\end{array}}
\newtheorem{theo}{Theorem}
\newtheorem{defi}{Definition}
\newtheorem{lem}{Lemma}

\newtheorem{propo}{Proposition}
\newtheorem{cor}{Corollary}

\title{Adaptive Quantum Computation, Constant Depth Quantum Circuits
and Arthur-Merlin Games}
\author{
Barbara M. Terhal\thanks{IBM Watson Research Center, {\tt
terhal@watson.ibm.com}. Partially supported by the National
Science Foundation under Grant. No. EIA-0086038 (as postdoctoral
scholar at the IQI, Caltech) and the National Security Agency and
the Advanced Research and Development Activity through Army
Research Office contract number DAAD19-01-C-0056.} \and David P.
DiVincenzo
\thanks{IBM Watson Research Center, {\tt divince@watson.ibm.com}.
Partially supported by National Security Agency, the Advanced
Research and Development Activity through Army Research Office
contract number DAAD19-01-C-0056 and the National Reconnaissance
Office.}}
\date{\today}
\begin{document}

\maketitle

\begin{abstract}
We present evidence that there exist quantum computations that can
be carried out in constant depth, using 2-qubit gates, that cannot
be simulated classically with high accuracy. We prove that if one
can simulate these circuits classically efficiently then ${\rm
BQP} \subseteq {\rm AM}$.
\end{abstract}


\section{Introduction}

The idea of quantum teleportation \cite{bennett+:tele},
transferring a quantum state by dual usage of classical
measurement data and quantum entanglement, has found a profound
application in quantum computation. It has been understood by
Gottesman and Chuang \cite{GC:tele} that not only states, but also
quantum gates, can be teleported. This observation has given rise
to a new paradigm for quantum computation, which we will refer to
as {\em adaptive} quantum computation.

In adaptive quantum computation, the outcomes of measurements
performed throughout the course of the computation determine the
quantum gates that are subsequently performed on the quantum
registers \footnote{Bernstein and Vazirani \cite{BV:qcomp} proved
that any model of quantum computation with measurements during the
computation is no more powerful than a model with measurements at
the end of the computation. Thus the rationale behind adaptive
quantum computation is merely that the set of quantum gates that
is needed to implement adaptive quantum computation may be smaller
and therefore simpler to implement {\em physically}.}; the quantum
program is ``adapted" to the classical measurement data.

{\em Nonadaptive} quantum computation, which we explore in this
paper, is a new computational model derived from the adaptive
scheme.  In this model we introduce a `guess' bit string $g$; $g$
is a prior guess of the outcome of all of the quantum measurements
performed in the course of the adaptive quantum computation.  This
guess can be compared with the actual quantum measurement outcomes
(which can now be deferred to the end of the computation, since no
quantum gate operations depend on them). In the (rare) case that
all these outcomes agree with the guess, the quantum computation
can be called successful, and {\em we know} when it occurs.
This suggests such quantum circuits could be hard to simulate
classically, i.e. if we were able to consider all possible
outcomes in our classical simulation, --including the rare one
that corresponds to having guessed correctly-- , then our
classical simulation would simulate the output of a
polynomial-time quantum computer on the remaining output bits.

On the other hand, the interesting feature of nonadaptive models
is that it can lead to quantum circuits of restricted depth.  The
adaptive model by Gottesman and Chuang (and similarly the one in
\cite{RB:oneway}) in which two-qubit gates are `teleported' into
the circuit and single-qubit gates are performed normally, maps to
a nonadaptive model of {\em constant} depth (Lemma \ref{prop2}).
That is, the resulting quantum circuit can be implemented in a
{\em constant} number of time steps, which seems to make it very
weak. The idea of computation by teleportation is also the basis
for the Knill-Laflamme-Milburn proposal of quantum computation by
linear optics \cite{KLM:lo} (see also the scheme by Gottesman,
Kitaev and Preskill \cite{GKP}). A nonadaptive version of the
Knill-Laflamme-Milburn construction also exists; however, because
this model, involving a quantum circuit with passive linear optics
applied to single photon states and followed by photon counting
measurements, does not conform to the standard qubit model, the
resulting nonadaptive circuit is not of constant depth.  It could
be of reduced depth, for example logarithmic depth, but this has
not been proved.

The fact that nonadaptive parallelized quantum models can be of
constant depth makes it likely (although not certain) that such
models have no (quantum) computational power whatsoever.  One may
also expect that they can be simulated efficiently by a
classical algorithm. However, we will provide evidence in this
paper that, in fact, it may be hard to simulate these models
classically. The results are the following:



1) We imagine that there is style of classical simulation that is
powerful enough to follow any computational pathway, regardless of
its probability of occurrence. We call this type of simulation by
density computation. At the end of the paper we prove that if such
a simulation of the constant depth quantum circuits is possible
then ${\rm BPP}={\rm BQP}$ and also the polynomial hierarchy
collapses (Corollary \ref{cor2}).

2) Even if the simulation does not have this extended power, there
is another well-known technique of endowing such a simulation with
greater power; if an all-powerful `Merlin' can direct the use of
the simulation by `Arthur', then Merlin can steer the simulation
to the rare, successful cases.  If such a simulation of constant
depth quantum circuits were possible, it would mean that
polynomial-time classical circuits, assisted by Merlin, would
necessarily be able to simulate any polynomial-time quantum
circuit. In complexity-theoretic language, this would imply the
containment ${\rm BQP} \subseteq {\rm AM}$ (${\rm
AM}$=`Arthur-Merlin') (Theorem \ref{theo2}).

\subsection{Previous work}
Constant-depth quantum circuits, but with gates that have {\em
arbitrary fan-in}, have been studied previously, see for example
Ref. \cite{green+}. The constant depth circuits that we consider
here have fan-in at most two.
In the setting of classical boolean circuits, there are arguments
that use a reduction to a constant-depth model by enumeration of
the inputs and outputs of each gate in the circuit \cite{hastad};
these arguments are somewhat akin to the ones we use here.



\section{Definitions}

\subsection{Basic Concepts}

We use the following standard notation. All strings are over the
alphabet $\Sigma=\{0,1\}$, $\Sigma^*$ is the set of all finite
strings and $|x|$ denotes the length of the string $x$.

\begin{defi}[Quantum Register $QR(w)$] A quantum register of $w$
qubits is a tensor product of 2 dimensional `qubit' Hilbert spaces
${\cal H}_2$, ${\cal H}={\cal H}_2^{\otimes w}$ of total dimension
$2^w$. Allowed states are unit-norm vectors $\ket{\psi}$ in this
space. The complex inner product in this space is denoted by
$\bra{v} w\rangle=\sum_i v_i^* w_i$ where $\ket{v}=\sum_i v_i
\ket{i}$ and $i$ is $w$-bit string. The standard (computational)
basis is formed by $w$-bit strings
$\ket{i}=\ket{i_1,i_2,i_3,...,i_w}$.
\end{defi}

\begin{defi}[Quantum Gate] The action of a general one-qubit gate is described by
an element of SU(2), which is applied to the vector describing the
state of that qubit.  The action of a general two-qubit gate is
described by an element of SU(4), which is applied to the vector
describing the state of that pair of qubits.  It is understood
that if a one- or two-qubit gate is applied to particular qubit(s)
of a register $QR(w)$, the unitary transformation on the full
$2^w$-dimensional state vector is obtained by a tensor product of
the gate action on the specified qubit(s) with the identity
operation on all other qubits.
\end{defi}

\begin{defi}[Quantum Measurement $\mathcal{M}$ in the Standard Basis]
A quantum measurement in the standard basis is an operation
applied to one qubit in a quantum register.  If the number of
qubits in the register is $w$, then the output of the measurement
is a single bit $b\in\Sigma$ and a smaller quantum register, of
size $w-1$, consisting of the unmeasured qubits.  The bit $b$
occurs with probability $p(b)=Tr(I\otimes\proj{b}\proj{\psi})$.
Here $I$ is the identity operator on the $2^{w-1}$-dimensional
Hilbert space of the unmeasured qubits. When the measurement
outputs the bit $b$, the new state of the unmeasured qubits is
$\ket{\psi_b}=\sum_{b'\in\Sigma^{w-1}}\ket{b'}\braket{b',b}{\psi}/\sqrt{p(b)}$.
\end{defi}

Measurements $\mathcal{M}$ on different qubits commute, so we may
consider a composite standard measurement on $k$ qubits with
output bit string $b$ as $k$ measurements applied to each of these
qubits. Two-qubit measurements
can be considered as a composite of some two-qubit quantum gate
applied to the pair of qubits to be measured, immediately followed
by a standard measurement $\mathcal{M}$ on the two qubits. Of
particular importance is the two-qubit measurement called a {\em
Bell measurement}.  For this, the two qubit unitary transformation
preceding ${\cal M}$ is of the form
$U=\ket{00}\bra{\Psi^+}+\ket{01}\bra{\Psi^-}
+\ket{10}\bra{\Phi^+}+\ket{11}\bra{\Phi^-}$, where the Bell states
are defined as
$\ket{\Phi^\pm}={1\over\sqrt{2}}(\ket{00}\pm\ket{11})$ and
$\ket{\Psi^\pm}={1\over\sqrt{2}}(\ket{01}\pm\ket{10})$.

\begin{defi}
[Quantum Circuit $QC_x(w,w',\ket{init},d)$] A quantum circuit,
determined by bit string $x$, of input width $w$, depth $d$, and
classical output width $w'\leq w$ consists of a quantum register
$QR(w)$ set initially (called time-step 0) to Hilbert-space vector
$\ket{init}$.  In each time-step 1 through $d-1$ the quantum
register is acted upon by one qubit or two qubit gates. Time-step
$d$ consists of a standard quantum measurement ${\cal M}$ applied
to some subset of $w'\leq w$ of the qubits; the output of the
quantum circuit is a bit string of length $w'$ and a $w-w'$-qubit
quantum state.
\end{defi}

We will often abbreviate our notation for a quantum circuit to
$QC(\ket{init},d)$.  Quantum circuits can be composed: \beq
QC_{x_2}(w-w',w'',\ket{\psi_b},d_2)*QC_{x_1}(w,w',\ket{init},d_1)\eeq
denotes a new circuit in which the unmeasured qubits of circuit
$QC_{x_1}$, in state $\ket{\psi_b}$ ($b$ is the bit string of the
measurement outcome), are fed into a second quantum circuit
$QC_{x_2}$. So long as $QC_{x_2}$ does not depend on $b$, this
composed circuit is identical to some single circuit
$QC'_x(w,w'',\ket{init},d')$ with $d'\leq d_1+d_2$. But in the
following we will consider cases where $QC_{x_2}$ {\em does}
depend on the measurement outcomes $b$.


\begin{defi}[{\rm BQP}] A language $L \subseteq \Sigma^*$ is in ${\rm BQP}$ if
$\forall x \in L$ $M(x)=1$ with probability larger than or equal
to $2/3$ where M is a uniformly generated family of quantum
circuits $QC_x(w,w',\ket{0},d)$ with w and d polynomial in $|x|$,
and one of the $w'$ output bits gives the value $M(x)$; when $x
\not\in L$ then $M(x)=1$ with probability smaller than or equal to
$1/3$.
\end{defi}

The class AM defined by Babai (\cite{babai} and \cite{BM:am}) is
an extension of the nondeterministic class NP where we allow
randomness and interaction in the verification procedure. Here is
a formal definition:

\begin{defi}[{\rm AM}] A language $L
\subseteq \Sigma^*$ is in AM if, $ \forall x \in L$ there exists a
strategy for Merlin such that a polynomial time computation in
$|x|$ by Arthur accepts with probability larger than or equal to
$2/3$. If $x\not \in L$ then for all strategies of Merlin, Arthur
accepts with probability smaller than $1/3$.  A `strategy' is
implemented in the following way: Arthur begins by sending a
random bit string $b$ ($|b|=poly(|x|)$)
to Merlin.  Merlin, after performing some (arbitrarily powerful)
classical computation on $b$, obtains bit string $m$ which he
returns to Arthur.  Arthur performs a polynomial time computation
with $m$ as input, obtaining a single-bit output $M(x)$.  The
acceptance probability is $p(M(x)=1)$.
\end{defi}

\subsection{Adaptive and nonadaptive quantum computation}


We now formalize what we mean by an {\em adaptive} quantum
computation model:

\begin{defi}[${\cal QC}_{ad}$] The class of adaptive quantum
computations ${\cal QC}_{ad}$ consists of all composed circuits of
the form \bea QC_{x,b_1,b_2,...b_R}(\ket{\psi_{b_R}},d_{R+1})*
...*
QC_{x,b_1,b_2}(\ket{\psi_{b_2}},d_3)*
QC_{x,b_1}(\ket{\psi_{b_1}},d_2)*
QC_x(\ket{init},d_1). \nonumber\eea This composition is adaptive
in the sense that the description of the second circuit
$QC_{x,b_1}$ is a function of the measurement outcome of the first
circuit $b_1$ as well as of the problem specification $x$; the
description of the third circuit $QC_{x,b_1,b_2}$ is a function of
the measurement outcomes of the first two, and so on. The depth
parameters $d_1$ through $d_{R+1}$, and the number of rounds
$R+1$, should be polynomial in $|x|$, as should all the widths;
also, for uniformity, the boolean functions determining the
quantum circuit at every round from the previous measurement
outcomes and the input $x$ should be efficiently
implementable.\label{nad}
\end{defi}

While ${\cal QC}_{ad}$ has no greater power than the set of
ordinary quantum circuits ${\cal QC}$, Gottesman and Chuang's work
shows that the universal operations for ${\cal QC}_{ad}$ could be
quite different than for ${\cal QC}$.  Their main theorem, stated
informally, is that one-qubit gates alone are sufficient to
implement all operations in ${\cal QC}_{ad}$, provided that
$\ket{init}$ consists of a sufficient supply of entangled states
as well as qubits in the $\ket{0}$ state, and that the quantum
circuits composing ${\cal QC}_{ad}$ are permitted to perform Bell
measurements (see Sec. \ref{GC+}).

Now we introduce a new model, nonadaptive quantum computation
which we derive from ${\cal QC}_{ad}$ and which we will use to
deduce interesting constraints relating simulatability of quantum
circuits and quantum complexity classes.


\begin{defi}[${\cal QC}_{nad}$] With each member $QC_{ad} \in {\cal
QC}_{ad}$ we associate a set of members $QC_{nad}(g) \in {\cal
QC}_{nad}$, one for each distinct value of the `guess' bit string
$g$. The nonadaptive quantum computation so obtained is the
composed quantum circuit \bea
QC_{x,g_1,g_2,...g_R}(\ket{\psi_{b_R}},d_{R+1})* ...
QC_{x,g_1,g_2}(\ket{\psi_{b_2}},d_3)
*QC_{x,g_1}(\ket{\psi_{b_1}},d_2)*
QC_x(\ket{init},d_1).
\eea
%
\end{defi}

The only difference between $QC_{ad}$ and $QC_{nad}(g)$ is that
the circuit's dependence on the measured values $b_1,b_2,...$ in
$QC_{ad}$ is replaced by the guessed values $g_1,g_2,...$ in
$QC_{nad}(g)$. Because of this, all intermediate measurements in
the nonadaptive circuit can be moved to the end, and the circuit
can be viewed as a single ordinary quantum circuit, with no
measurements during the computation.


\section{Classical Simulations}

We formalize the notion of classical simulatability with a certain
accuracy:

\begin{defi} [$S_\epsilon({\cal QC})$] An efficient
simulation $S_\epsilon({\cal QC})$ (with accuracy parameter
$\epsilon$) of a (uniformly generated) family of quantum circuits
${\cal QC}$ exists if for each $QC(w,w',\ket{0},d) \in {\cal QC}$
there is a classical boolean circuit with depth $d'$ and input
width $r$ ($r,d'=poly(w,d)$), and output width $w'$, such that
$\forall b$, $|N(b)/2^r-p_{QC}(b)|\leq \epsilon p_{QC}(b)$. Here,
$p_{QC}(b)$ is the probability that the measured state of the
output quantum register is $b$ for a particular quantum circuit
$QC$; $N(b)$ is the number of settings of the input register of
the classical circuit that simulates QC, for which the classical
circuit outputs $b$. The classical circuits are uniformly
generated from the description of the $QC$-circuits. \label{epsac}
\end{defi}

A stronger type of classical simulation is one where one can
explicitly calculate the (conditional) probability of a certain
set of outcomes, and then sample this probability distribution.
Here is our definition:


\begin{defi}[$S^C({\cal QC})$] An efficient
density computation of a (uniformly generated) family of quantum
circuits ${\cal QC}$ is one that proceeds as follows: we first
divide up the full quantum measurement at the end of $QC \in {\cal
QC}$ into separate measurements $\mathcal{M}_1$, $\mathcal{M}_2$,
... on disjoint sets of qubits that contain a constant number of
qubits. Let $b_i$ denote the set of potential outcomes of
measurement $M_i$, and let ${\sf b_i}$ denote an outcome.
An efficient density computation exists if there exist
polynomial-time (in the width and depth of $QC(w,w',\ket{0},d) \in
{\cal QC}$) uniformly generated classical procedures for
evaluating
the conditional probabilities \beq p(b_i|{\sf b_{j_1}} \ldots {\sf
b_{j_k}}). \label{condprob} \eeq Here the set of indices
$j_1,\ldots,j_k$ can be any subset (including the empty set) of
the set of measurements ${\cal
M}=\{\mathcal{M}_1,\mathcal{M}_2,...\}$ and $i \neq j_1 \ldots
j_k$ labels any other measurement.
\end{defi}

For an example of a density computation, see the simulation
algorithm in Ref. \cite{valiant:simulation_siam}. This definition
leads obviously to

\begin{propo} If an efficient density computation exists, it
provides the means for performing an efficient simulation with
$\epsilon=0$.
\end{propo}

\begin{proof}
 We proceed as follows. We pick a first measurement ${\cal M}_1$ (it
does not matter which one since they all commute) and calculate
$p(b_1)$, i.e. we calculate a constant number $c_1$ of
probabilities, where $c_1$ is the number of outcomes of ${\cal
M}_1$. We flip coins according to the probability distribution
$p(b_1)$ and fix the outcome ${\sf b_1}$. Then we pick a next
measurement ${\cal M}_2$ and we calculate $p(b_2|{\sf b_1})$,
again a constant number of probabilities. We flip coins, fix the
outcome and proceed to the next measurement ${\cal M}_3$ etc.
There are no more than $w$ measurements (for a quantum circuit
$QC$ of width $w$) and thus in total we calculate at most $w\max_i
c_i$ conditional probabilities.
\end{proof}

{\em Remark}: One may also consider density computations with
accuracy parameter $\delta$ in which the conditional probabilities
can be calculated as $|p_{sim}(.|.)-p(.|.)| \leq \delta p(.|.)$.
If $\delta=2^{{1\over w}\log(1+\epsilon)}-1$ where $w$ is the
width of the quantum circuit (for large width $w$, $\delta
\rightarrow {1\over w}\ln(1+\epsilon)$), then this will provide a
way to do an simulation with accuracy parameter $\epsilon$.

Definitions 9 and 10 relate to regular quantum circuits ${\cal
QC}$ but can be extended in a straightforward way to adaptive
quantum circuits ${\cal QC}_{ad}$ by restricting ourselves in the
density computation Definition 10 to estimating conditional
probabilities where (intermediate) measurement ${\cal M}_i$ occurs
{\em after} the (intermediate) measurements labeled by $j_1,
\ldots, j_k$.

In the `density computation' setting, there is a close connection
between the simulation of adaptive and nonadaptive circuits:

\begin{theo}
If there exists an efficient density computation of $QC_{nad}(g)
\in {\cal QC}_{nad}$ for all $g$, then there is a simulation with
accuracy parameter $(\epsilon=0)$ for $QC_{ad} \in {\cal
QC}_{ad}$. \label{theo1}
\end{theo}

\begin{proof} The simulation of $QC_{ad}$ will be a
direct adaptation of the algorithm for the density computation of
${QC}_{nad}(g)$ for some $g$.  We consider the composed quantum
circuit in Definition \ref{nad} for $QC_{ad}$.  The first member
of this composition, $QC_x(\ket{init},d_1)$, is by itself an
example of a nonadaptive circuit and can be simulated efficiently
by the hypothesis. We flip coins biased according to the outcome
probabilities of ${\cal M}_1$ (the measurement following $QC_x$)
and we fix the outcome, say $b_1={\mathsf b}_1$. This fixes the
second circuit in the composition $QC_{x,{\mathsf b}_1}$ and any
further choices of gates and measurements depending on $b_1$. Now
we consider two stages of the quantum circuit, $QC_{x,{\mathsf
b}_1}(\ket{\psi_{{\mathsf b}_1}},d_2)*QC_x(\ket{init},d_1)$. Since
the second circuit is now fixed, it is no longer adaptive and we
may consider the pair as a single nonadaptive quantum circuit with
guess bit $g={\sf b}_1$; the measurements ${\cal M}_1$ and ${\cal
M}_2$ (the measurement following $QC_{x,b1}$) may be moved to the
end. Since we have already fixed the outcome of measurement ${\cal
M}_1$, we are interested in sampling from the probability
distribution of outcomes of ${\cal M}_2$ given $b_1={\mathsf
b}_1$. By hypothesis, our classical algorithm can do this by, for
example, letting us calculate the conditional probability
distributions $p(b_2({\cal M}_2)|{\mathsf b}_1)$. By computing the
constant number of quantities $p(b_2({\cal M}_2)|{\mathsf b}_1)$
and flipping coins we can implement measurement ${\cal M}_2$. We
fix its outcome, say $b_2={\mathsf b}_2$ and proceed as before to
the third measurement by moving it to the end and using the
density computation of some nonadaptive circuit with guess bits
$g={\sf b}_1 {\sf b}_2$, etc. Note that if ${\cal M}_1$, ${\cal
M}_2$, etc., do not have a constant number of outcomes, the
circuit can always be broken up into a larger number of stages,
where the number of outcomes for each of these sub-stages is
constant.
\end{proof}

A special case occurs when the probability distribution over the
output bits $b_1, \ldots, b_k$ of a circuit $QC_{ad}$ does not
depend on the outcomes of the intermediate measurements, in other
words $QC_{ad}$ represents the same logical circuit on the qubits
{\em independent of the outcome of these intermediate
measurements}. An example is the Gottesman-Chuang construction
considered in the next section. These kind of circuits are the
ones of interest since we want to implement a fixed circuit and
not a mixture of circuits. We denote this class of circuits with
an additional label `fix', i.e. ${\cal QC}_{ad,fix}$. In that case
we also have

\begin{theo}
If there exists an efficient density computation of
$QC_{nad,fix}(g) \in {\cal QC}_{nad,fix}$ for all $g$, then there
is an efficient density computation for $QC_{ad,fix} \in {\cal
QC}_{ad,fix}$. \label{theo2}
\end{theo}

\begin{proof}
Consider $QC_{ad,fix}$ with a certain set of intermediate
measurement outcomes $g'$. We know that the probability
distribution of outcomes of $QC_{ad,fix}$ $p(b_1 \ldots b_k|g')$
is identical to the probability distribution of outcomes of
$QC_{nad,fix}(g')$. Since also $p(b_1 \ldots b_k|g')=p(b_1 \ldots
b_k)$ the efficient density computation of $QC_{nad,fix}(g')$ can
be directly used to do an efficient density computation of
$QC_{ad,fix}$.
\end{proof}

Our second theorem has interesting consequences when applied to a
subset of ${\cal QC}_{ad,fix}$ circuits that give universal
quantum computation:

\begin{cor} Suppose the set $\{QC^U_{ad,fix}\}\subset {\cal QC}_{ad,fix}$ contains a
universal set of quantum circuits, sufficient to implement all
polynomial-time quantum computations (where each $QC^U_{ad,fix}$
is such that the logical circuit on the qubits does not depend on
the outcomes of the intermediate measurements). If there is an
efficient density computation of the corresponding nonadaptive set
$\{QC^U_{nad,fix}(g)\}$ for all $g$, then for the polynomial
hierarchy ${\rm PH}$ we have
 ${\rm PH}={\rm BPP}={\rm BQP}$ (thus the polynomial
hierarchy would collapse to $\Sigma_2^P$ since ${\rm BPP} \in
\Sigma_2^P$ \cite{lautemann}).\label{cor1}
\end{cor}

\begin{proof}
If the simulations of the nonadaptive circuits are possible, then
by Theorem \ref{theo1} all the adaptive simulations are also
possible. The density computation of Theorem \ref{theo2} does more
than simply providing a simulation; it provides a means of
calculating the outcome probability of any polynomial depth
quantum circuit. It has been shown \cite{fenner+:ph} that
determining the acceptance probability of a quantum computation
which we would be able to do if we could calculate all joint
probabilities, is equivalent to the complexity class ${\rm
coC\!\!=\!\!P}$. Therefore we would have $ {\rm coC\!\!=\!\!P}
\subseteq {\rm BPP}$. On the other hand, it is known that ${\rm
PH} \subseteq {\rm BPP}^{\rm coC\!=\!P}$ and thus ${\rm PH}
\subseteq {\rm BPP}^{\rm BPP}={\rm BPP}$.
\end{proof}

This corollary has more explicit consequences if we consider its
application to the Gottesman-Chuang adaptive computation model,
which we now examine in more detail.

\section{Constant-Depth Quantum Circuits}
\label{GC+}

\begin{figure}[h]
\epsfxsize=14cm \epsffile{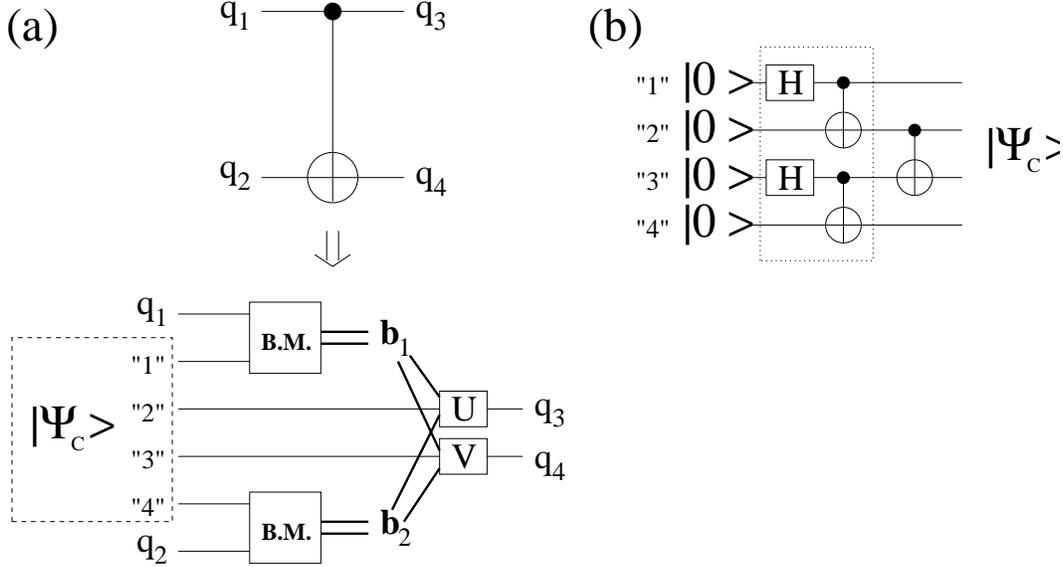} \caption{(a) The
Gottesman-Chuang implementation of the CNOT gate by teleportation.
In addition to the two qubit inputs $q_1$ and $q_2$, the
teleportation circuit has four additional ancilla qubit inputs
(``1", ``2", ``3", ``4") preset to the entangled state
$\ket{\Psi_C}$. Two Bell measurements (B.M.) are performed as
indicated, resulting in two bit-pairs ${\bf b}_1$ and ${\bf b}_2$
as output. These bit pairs determine the parameters of two
one-qubit quantum gates $U$ and $V$. (b) Construction of the
entangled state $\ket{\Psi_C}$. $H$ is the one-qubit Hadamard
gate, specified by the $2\times 2$ matrix ${1\,\,1\choose 1
-1}/\sqrt{2}$.  The dotted box (which can be completed in one time
step) causes the creation of the entangled state $\Phi^+$ between
ancilla bits ``1" and ``2" and between ``3" and ``4". The final
CNOT constitutes the ``offline" application of the two-qubit gate
as mentioned in the text.} \label{figy}
\end{figure}

The Gottesman-Chuang construction for quantum computation starts
from the well-known fact that there exist universal quantum gate
sets containing only a single two-qubit gate, the controlled-NOT,
along with certain one-qubit gates.
They obtain adaptive circuits from such a standard quantum circuit
by a one-for-one replacement of each CNOT in the circuit by the
teleportation protocol outlined in Fig. \ref{figy}(a).  The
entangled four-qubit state $\ket{\Psi_C}$ can be created
``offline" at the beginning of the computation by the procedure
shown in Fig. \ref{figy}(b).

We formally define the class of circuits ${\cal
GC}_{ad}\subset{\cal QC}_{ad}$ obtained in this fashion:

\begin{defi}[${\cal GC}_{ad}$] A quantum circuit $G\in {\cal GC}_{ad} \subset {\cal QC}_{ad}$ when
1) All the quantum circuits $QC_{b_1,b_2,...}$ composing $G$
contain only one-qubit gates; 2) The initial state $\ket{init}$
consists of some qubits in state $\ket{0}$ and others in the
entangled four qubit state $\ket{\Psi_C}$. The qubits of $\Psi_C$
are labeled ``1" to ``4", see Fig. \ref{figy}(b); 3) Each
intermediate measurement consists of two Bell measurements. The
first of these Bell measurements (see Fig. \ref{figy}(a)) uses
qubit ``1" of a `new' $\Psi_C$ that has not been acted upon
previously, and some other arbitrary qubit in the circuit. The
second measurement uses qubit ``4" of the {\em same} $\Psi_C$ and
some other arbitrary qubit.
\end{defi}


The nonadaptive circuits corresponding to ${\cal GC}_{ad}$ are of
constant depth:

\begin{lem} Consider any circuit $G\in {\cal GC}_{ad}$, and consider
the corresponding set of nonadaptive circuits $G(g)$ which
together make up the set ${\cal GC}_{nad}$.  Any circuit in ${\cal
GC}_{nad}$ has at most depth four, using one and two-qubit gates
and starting with the state $\ket{00 \ldots 0}$.\label{prop2}
\end{lem}

\begin{proof}
The essential idea behind the depth-reduction is that in the
nonadaptive model the Bell measurements can be done at the end and
all further gates in the quantum circuits are one-qubit gates
given a preparation of the states $\Psi_C$ (which can also be done
in a few time-steps). The nonadaptive circuit has the following
form. In the first time step $(I)$ one creates a set of entangled
states $\Phi^+$ by performing a set of 2-qubit gates all acting in
parallel. At the same time, we perform some single-qubit gates on
the other qubits, the `data qubits'. The second time step ({\em
II}) is built out of two steps. We do CNOT gates on halves of the
entangled states; the target and control bits are each half of a
different entangled state (so as to create $\Psi_C$). Then we do
some single-qubit gates on the qubits ``2" and ``3" (see Fig.
\ref{figy}(b)) that went through these CNOT gates. These two steps
can be merged into one by permitting arbitrary 2-qubit gates on
halves of the entangled states. Then we rotate some qubits from
the Bell basis to the standard basis ({\em III}) and in the final
time step we perform single-qubit measurements in the
computational basis ({\em IV}).
\end{proof}

The restricted form of the ${\cal GC}_{nad}$ circuits leads to a
conclusion about the simulatability of constant-depth quantum
circuits:

\begin{cor}
If there is an efficient density computation of all quantum
circuits $QC(\ket{0},d=4)$, then ${\rm PH} \subseteq {\rm BPP}$
and ${\rm BQP} \subseteq {\rm BPP}$ and thus ${\rm PH}={\rm
BPP}={\rm BQP}$.\label{cor2}
\end{cor}

\begin{proof} Immediate, by application of Corollary \ref{cor1}
and Lemma \ref{prop2}.
\end{proof}

This corollary indicates that a classical simulation, at least of
the density computation type, of constant depth quantum circuits,
is quite unlikely to be possible. This may be considered a
surprising result, since it is clear that for a quantum circuit of
constant depth $d$ and width $n$, any measurement on $\log n$
qubits can be efficiently reproduced by a classical density
computation, that is, the family of quantum circuits $\{QC(n,\log
n,\ket{0},d)\}$ is simulatable for constant $d$. In the past
history of a single qubit there is only a total constant number
($2^d$) of qubits with which it has interacted and therefore the
output state of $\log n$ qubits can be determined by following the
evolution of at most $2^d \log n$ input qubits. We can thus
simulate the measurements on $\log n$ qubits by following the
$n^{2^d}$ amplitudes during the computation and tracing over
outputs that are no longer in the past history cone of the output
qubits of interest.


We now note that the depth of four, singled out by the
Gottesman-Chuang construction, is the smallest depth for which
Corollary \ref{cor2} holds:

\begin{propo}
An efficient density computation exists of all quantum circuits of
depth three, $QC(\ket{0},d=3)$.
\end{propo}

\begin{proof}
We give the simulation.  After the first time step the quantum
state of the circuit consists of a set of 2-qubit entangled states
and possibly some 1-qubit states and thus the amplitudes of this
state can be efficiently represented classically. We may consider
the second computing step and the final measurement step as one
single step in which a set of final measurements are performed in
arbitrary 2-qubit bases. We pick a first measurement. It is simple
to calculate the probabilities for the various outcomes since they
depend on the state of no more than four qubits. We flip coins
according to the outcome probabilities and fix the outcome. We
replace the 4-qubit state by the post-measurement projected state
consisting now of 2 qubits. We pick the next measurement and
proceed similarly etc. If only a subset of these measured bits are
required as output, the rest are simply discarded.
\end{proof}

It is clear that adding a fourth layer of computation will break
the method of proof for the proposition. The problem is that each
final measurement in the Bell basis leaves an entangled state with
more qubits which may not have an efficient classical
representation. The reason that a constant-depth quantum
computation could be hard to simulate classically may be precisely
this.

\subsection{General Simulations}

As we argued above, the constant-depth ${\cal GC}_{nad}$ model is
able to perform quantum computation -- the probability of success
is exponentially small, but there is a flag that indicates when
the successful outcome is achieved.  This means that if there were
a classical algorithm of any sort that simulates the
constant-depth ${\cal GC}_{nad}$ model, then there would exist an
efficient classical probabilistic algorithm that could simulate a
polynomial-time quantum computation with exponentially small
probability, but with a success flag.  Even though such a
simulation would not be useful, it would nevertheless have some
interesting consequences for computational complexity classes. In
particular, we will relate the class ${\rm BQP}$ to the classical
nondeterministic complexity class AM (for Arthur-Merlin). We
consider the circuits $G(g=0) \in {\cal GC}_{nad}$ which have
guessed outcome corresponding to $U=I$ and $V=I$ after every Bell
measurement \cite{GC:tele}. The following theorem is proved for
these circuits with $g=0$, but holds equally well for other values
of $g$:

\begin{theo}
If an efficient simulation with accuracy parameter $\epsilon <
1/3$ exists for the family of circuits $\{G(g=0)\}\subset {\cal
GC}_{nad}$, then ${\rm BQP} \subseteq {\rm AM}$. \label{theo3}
\end{theo}

The idea behind the proof is the following.  The classical
probabilistic simulation of $G(g=0) \in {\cal GC}_{nad}$ uses a
certain number of random coins. For some set of values of these
coins the simulation outputs (1) the Bell measurement outcomes
corresponding to the guess string $g=0$ such that we know that the
simulated circuit performs a successful simulation of some quantum
computation $M$ and (2) the bit value 1 as the outcome of this
quantum computation $M$. Thus the size of this set of coin values
depends on whether $M$ outputs 1 with large probability or small
probability, corresponding to the decision problem that $M$
solves. The estimation of the approximate size of a set is a
problem that is known to be in {\rm AM}.

Here are the details of the proof:
\begin{proof}
Let $L \in {\rm BQP}$ and let $\{G_x\}\in {\cal GC}_{ad}$ be the
uniformly generated family of quantum circuits, of the
Gottesman-Chuang type, that output a bit $M(x)$ that decides $L$;
thus if $x\in L$, $M(x)=1$ with probability at least 2/3, while if
$x\notin L$, $M(x)=1$ with probability less than 1/3. Consider the
corresponding nonadaptive version of these $GC$ circuits
$G_{x}(g=0)$ where $g$ is the guess bit string ($|g|=k$) for all
the Bell measurements.  We will be interested in two of the
outputs of the circuit $G_{x}(g=0)$: The bit string $y$ giving all
the Bell measurement outcomes, and the decision bit $M'(x)$ where
$M'$ is a function that coincides with $M$ when $y=g=0$, and does
not, in general, coincide with $M$ when $y\neq 0$. Since all Bell
measurement outcomes are equally likely (see Appendix A), that is,
$p(y)=1/2^k$ for all $y$, then if $x\in L$, $p(y=g,M(x)=1)\geq
(2/3) \times (1/2^k)$.

The classical probabilistic simulation $S_{\epsilon}$ of the
circuit $G_{x}(g=0)$ takes as input a set of $n$ random bits $r$
($n=poly(|x|)$), the description of the circuit $G_x(g=0)$ (note
that it is necessary that $n\geq|g|=k$), and the input $x$. The
machine $S_{\epsilon}$ outputs ($y,M'(x))$ in $poly(n)$ time;
thus, we can express the input-output relation of the simulation
as $(r, G_x(g=0),x)\stackrel{S_\epsilon}{\rightarrow}(y,M'(x))$.
Since $S_{\epsilon}$ is a good simulation of $G_{x}$, there exist
a set of values for the random bits $r$ such that the Bell
outcomes agree with the guesses, $(r,
G_x(g=0),x)\stackrel{S_\epsilon}{\rightarrow}(y=0,M(x))$. In fact,
according to the accuracy parameter $\epsilon$ condition of
Definition \ref{epsac} we have that, for this simulation,
$p(y=0,M(x)=1)_{sim}\geq (1-\epsilon)(2/3) \times 2^{-k}$. Since
there are $2^n$ strings $r$, the total size of this set ${\bf S}$
of random coin settings for which $y=0$ and $M(x)=1$ is at least
${\sf BIG}=(1-\epsilon)(2/3) \times 2^{n-k}$; if $x\notin L$, then
the size of the set ${\bf S}$ for which $y=0$ and $M(x)=1$ is
guaranteed to be less than or equal to ${\sf
SMALL}=(1+\epsilon)(1/3) \times 2^{n-k}$.

Thus, if the simulation $S_\epsilon$ exists, then the problem of
deciding whether input $x$ is in a BQP language L or not is
equivalent to determining whether this set ${\bf S}$ of $n$-bit
strings is larger than or equal to size ${\sf BIG}$ or smaller
than or equal to ${\sf SMALL}$, where membership in the set is
easy to determine (in polynomial time) by running the simulation
$S_\epsilon$.  If $\epsilon<1/3$, then we are guaranteed that
${\sf BIG} >{\sf SMALL}$.  This problem of determining
``approximate set size'' is known to be solvable as a two-round AM
game (see Lemma 1 in \cite{GS:coin}). In Appendix B we explain how
the game proceeds. \end{proof}

How unlikely is the containment BQP $\subseteq$ AM?  Nothing is
definitely known, but the consensus is that it is rather unlikely.

\section{Conclusion}

The nonadaptive Gottesman-Chuang circuit is a very curious
resource. According to the evidence given by this paper, its
multiple-bit output is hard to generate classically. Still, it is
an open question whether, for example, a class such as ${\rm
BPP}^{{\cal GC}_{nad}}$ would have additional power over ${\rm
BPP}$ (see for example the recent results in Ref.
\cite{fenner+:condep}).

From an experimental physics point of view it is clear that it
would be extremely interesting to find a problem in ${\rm
BPP}^{{\cal GC}_{nad}}$ which is not known to be in ${\rm BPP}$; a
constant-depth quantum circuit should be easier to build than a
universal quantum computer.

Our Theorem 1 holds for any adaptive quantum computation model
with its corresponding nonadaptive version, including, for
example, the Knill-Laflamme-Milburn (KLM) scheme.  So, our results
can be viewed as evidence that a nonadaptive KLM scheme, i.e. mere
linear optics on Fock states followed by photon counting
measurements, may perform some interesting non-classical
computation.

\subsection*{Acknowledgments}

We would like to thank Manny Knill for a stimulating email
correspondence on the various quantum computation models. We would
like to thank Yaoyun Shi and Scott Aaronson for helpful
discussions and comments concerning some computer science aspects
of this research. We thank Alexei Kitaev for his explanation of
the AM implementation of the approximate set size problem. We also
thank Aram Harrow for pointing out an inaccuracy in the original
version of this paper.

\bibliographystyle{abbrv}
\bibliography{refs}  

\newcommand{\etalchar}[1]{$^{#1}$}
\begin{thebibliography}{GHMP02}

\bibitem[Bab85]{babai}
L.~Babai.
\newblock Trading group theory for randomness.
\newblock In {\em Proceedings of 17th STOC}, pages 421--429, 1985.

\bibitem[BBC{\etalchar{+}}93]{bennett+:tele}
C.H. Bennett, G.~Brassard, C.~Cr{\'e}peau, R.~Jozsa, A.~Peres, and W.K.
  Wootters.
\newblock Teleporting an unknown quantum state via dual classical and
  {Einstein-Podolsky-Rosen} channels.
\newblock {\em Phys. Rev. Lett.}, 70:1895--1899, 1993.

\bibitem[BM88]{BM:am}
L.~Babai and S.~Moran.
\newblock Arthur-{M}erlin games: a randomized proof system and a hierarchy of
  complexity classes.
\newblock {\em Journal of Computer and System Sciences}, 36:254--276, 1988.

\bibitem[BV97]{BV:qcomp}
E.~Bernstein and U.~Vazirani.
\newblock Quantum complexity theory.
\newblock {\em SIAM Journal on Computing}, 26(5):1411--1473, 1997.

\bibitem[CW79]{CW:hash}
J.~L. Carter and M.~N. Wegman.
\newblock Universal classes of hash functions.
\newblock {\em Journal of Computer and Systems Sciences}, 18(2):143--154, 1979.

\bibitem[FGHP99]{fenner+:ph}
Stephen~A. Fenner, Frederic Green, Steven Homer, and Randall Pruim.
\newblock Determining acceptance possibility for a quantum computation is hard
  for the polynomial hierarchy.
\newblock {\em Proc. Roy. Soc. London A}, 455:3953--3966, 1999.

\bibitem[FGHZ03]{fenner+:condep}
S.~Fenner, F.~Green, S.~Homer, and Y.~Zhang.
\newblock Bounds on the power of constant-depth quantum circuits.
\newblock Preprint quant-ph/0312209 December 2003.

\bibitem[GC99]{GC:tele}
D.~Gottesman and I.~Chuang.
\newblock Demonstrating the viability of universal quantum computation using
  teleportation and single qubit operations.
\newblock {\em Nature}, 402(6760):390--393, 1999.

\bibitem[GHMP02]{green+}
Frederic Green, Steven Homer, Christopher Moore, and Christopher Pollett.
\newblock Counting, fanout, and the complexity of quantum {ACC}.
\newblock {\em Quantum Information and Computation}, 2(1):35--65, 2002.

\bibitem[GKP01]{GKP}
D.~Gottesman, A.Yu. Kitaev, and J.~Preskill.
\newblock Encoding a qubit in an oscillator.
\newblock {\em Phys. Rev. A}, 64:012310, 2001.

\bibitem[GS86]{GS:coin}
S.~Goldwasser and M.~Sipser.
\newblock Private coins versus public coins in interactive proof systems.
\newblock In {\em Proceedings of 18th STOC}, pages 59--68, 1986.

\bibitem[GS89]{GS:coin_book}
S.~Goldwasser and M.~Sipser.
\newblock Private coins versus public coins in interactive proof systems.
\newblock {\em Advances in Computing Research: a research annual}, 5:73--90,
  1989.

\bibitem[H{\aa}s87]{hastad}
J.~H{\aa}stad.
\newblock One-way permutations in {NC}$^0$.
\newblock {\em Information Processing Letters}, 26:153--155, 1987.

\bibitem[KLM01]{KLM:lo}
E.~Knill, R.~Laflamme, and G.~Milburn.
\newblock A scheme for efficient quantum computation with linear optics.
\newblock {\em Nature}, 409:46--52, 2001.

\bibitem[Lau83]{lautemann}
C.~Lautemann.
\newblock {BPP} and the polynomial hierarchy.
\newblock {\em Information Processing Letters}, 17(4):215--217, 1983.

\bibitem[RB01]{RB:oneway}
R.~Raussendorf and H.~Briegel.
\newblock A one-way quantum computer.
\newblock {\em Phys. Rev. Lett.}, 86:5188--5191, 2001.

\bibitem[Val02]{valiant:simulation_siam}
L.~Valiant.
\newblock Quantum circuits that can be simulated classically in polynomial
  time.
\newblock {\em SIAM J. Comput.}, 31(4):1229--1254, 2002.

\end{thebibliography}
%
%

\appendix

\section{Appendix: Bell outcomes}

In the Gottesman-Chuang circuit (adaptive or nonadaptive), each
Bell measurement is performed on uncorrelated qubits, one of which
is a qubit ``1" or qubit ``4" of the state $\Psi_C$. Thus, the
density matrix of the two-qubit system has the form \beq
\rho_{AB}={1\over 2}I\otimes\rho_B \eeq for some choice of the AB
labels.  The probability that the Bell measurement outcome is,
say, $\Phi^+$, is by standard quantum mechanical rules
$\bra{\Phi^+}\rho_{AB}\ket{\Phi^+}=$
\beq
%
{1\over
4}(\bra{00}I\otimes\rho_B\ket{00}+\bra{00}I\otimes\rho_B\ket{11}+
\bra{11}I\otimes\rho_B\ket{00}+\bra{11}I\otimes\rho_B\ket{11})=
{1\over 4}{\rm Tr} \rho_B={1\over 4},\eeq and similarly for the
other three Bell states.  Thus the probability of outputting any
bit pair is uniform ($p=1/4$) as claimed.

\section{Appendix: Approximate Set Size}

We begin with a guaranteed separation of set sizes, ${\sf
BIG}/{\sf SMALL}\geq \frac{2-2\epsilon}{1+\epsilon}$. The first
step is to amplify this ratio to a larger number, say ${\sf
BIG}/{\sf SMALL}=d/8$ (we follow here the notation of the lemma in
\cite{GS:coin}), by considering
$u=\log(d/8)/\log(\frac{2-2\epsilon}{1+\epsilon})$ runs of the
simulation; that is, we consider the new set ${\bf
S}'=\underbrace{{\bf S} \times {\bf S} \times \ldots \times {\bf
S}}_u$. Given that ${\bf S}'$ is a subset of all $u\times n$-bit
strings, and $p=\lfloor 8 \log {\sf BIG}\rfloor$ (i.e. $p\sim
u(n-k)$), the game proceeds as follows:
\begin{itemize}
\item Arthur, the verifier, picks at random $l=p+1$ hash functions
$h_1$,...,$h_l$, $h:\Sigma^{un}\rightarrow \Sigma^p$ and $l^2$
random bit strings $Z=\{z_1,...,z_{l^2}\}$, $z_i\in \Sigma^p$; all
these are sent to Merlin, the prover.  The hash may be of the
Carter-Wegman type \cite{CW:hash}, so each hash function is
specified by a $ u n\times p$ random Boolean matrix. \item Merlin
attempts to respond with $t\in {\bf S}'$ such that, for some $i$,
$h_i(t)\in Z$.
\end{itemize}

With a suitably chosen amplification factor, the game can succeed
with almost certainty (but not absolute certainty); that is, if
$x\in L$, according to \cite{GS:coin} (see also
\cite{GS:coin_book}, the probability that Merlin can supply a
proof if $|{\bf S}'| \geq {\sf BIG}$ is at least as large as
$1-2^{-l/8}$, while if $|{\bf S}'|\leq{\sf SMALL}$ the probability
that Merlin can give a proof is no greater than $l^3/d$.  Thus, we
can make the failure probability in both directions exponentially
small in $|x|$ (an exponentially large $d$ only requires a
polynomially large number of repetitions $u$).


\end{document}